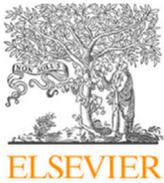
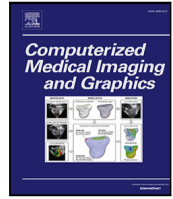

# Integrating features from lymph node stations for metastatic lymph node detection

Chaoyi Wu [a,1], Feng Chang [a,1], Xiao Su [c], Zhihan Wu [d], Yanfeng Wang [a,b], Ling Zhu [c,*], Ya Zhang [a,b,*]

[a] *Cooperative Medianet Innovation Center, Shanghai Jiao Tong University, Shanghai 200240, China*
[b] *Shanghai AI Laboratory, Shanghai 200232, China*
[c] *Department of Radiology, School of Medicine, Shanghai Ninth People's Hospital, Shanghai Jiao Tong University, Shanghai 200011, China*
[d] *School of Medicine, Shanghai Jiao Tong University, Shanghai 200025, China*

## ARTICLE INFO

## ABSTRACT

Metastasis on lymph nodes (LNs), the most common way of spread for primary tumor cells, is a sign of increased mortality. However, metastatic LNs are time-consuming and challenging to detect even for professional radiologists due to their small sizes, high sparsity, and ambiguity in appearance. It is desired to leverage recent development in deep learning to automatically detect metastatic LNs. Besides a two-stage detection network, we here introduce an additional branch to leverage information about LN stations, an important reference for radiologists during metastatic LN diagnosis, as supplementary information for metastatic LN detection. The branch targets to solve a closely related task on the LN station level, *i.e.*, classifying whether an LN station contains metastatic LN or not, so as to learn representations for LN stations. Considering that a metastatic LN station is expected to significantly affect the nearby ones, a GCN-based structure is adopted by the branch to model the relationship among different LN stations. At the classification stage of metastatic LN detection, the above learned LN station features, as well as the features reflecting the distance between the LN candidate and the LN stations, are integrated with the LN features. We validate our method on a dataset containing 114 intravenous contrast-enhanced Computed Tomography (CT) images of oral squamous cell carcinoma (OSCC) patients and show that it outperforms several state-of-the-art methods on the mFROC, maxF1, and AUC scores, respectively.

## 1. Introduction

Metastatic lymph nodes (LNs) play a pivotal part in the spread of original tumors. Detecting metastatic LNs in time is of great importance both in cancer prevention and treatment planning. However, metastatic LN detection is extremely challenging and time-consuming even for an experienced radiologist due to the following characteristics. Firstly, the sizes of metastatic LNs are extremely small compared to other organs and tissues. As shown in Fig. 1(a), most cases have a diameter of less than 20 mm. Secondly, metastatic LNs are usually sparsely distributed, with mostly 1–2 metastatic LNs per patient in our task (see Fig. 1(b)). Thirdly, the appearance of metastatic LNs is ambiguous, *i.e.*, the intra-class differences of metastatic LNs are large while the inter-class differences between metastatic LNs and normal LNs are small, as shown in Fig. 2. Radiologists must judge whether an LN candidate is metastatic by simultaneously examining its texture, shape, size, as well as its neighboring tissues and organs.

Several previous studies start with an easier task of detecting enlarged LNs (Barbu et al., 2011; Bouget et al., 2019; Nogues et al., 2016; Oda et al., 2018). The enlarged LNs are highly likely to be metastatic, but not vice versa. Such methods are thus limited in coverage as nearly 70% of metastatic LNs are within 10 mm in size, as shown in Fig. 1(a). Some subsequent studies consider small metastatic LNs and leverage PET modality or tumor location to provide additional information to assist the detection of metastatic LNs and achieve promising performance (Chao et al., 2020; Zhu et al., 2020a,c). However, in practice, patients do not always have PET modality available due to its high cost, or the primary tumors because they may have been surgically removed, which makes the above methods inapplicable. This paper thus explores leveraging other more accessible auxiliary information to help detect metastatic LNs, as an important supplement to the above methods.

In this paper, we propose an efficient deep learning based method to leverage information about LN stations to detect metastatic LNs. The LN

---

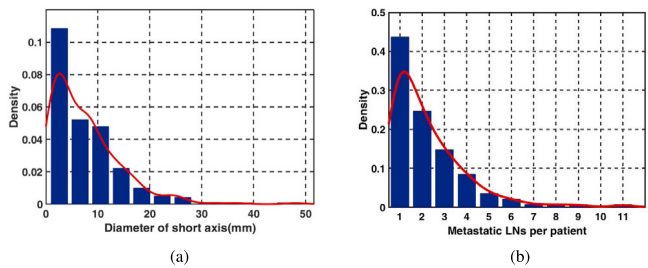

**Fig. 1.** Statistics of metastatic LNs. (a) Distribution of metastatic LNs sizes (short axis length). (b) Distribution of the number of metastatic LNs per patient.

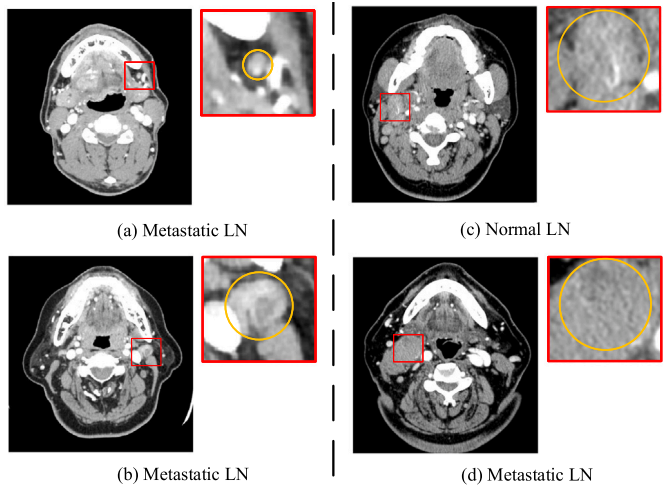

**Fig. 2.** Left two metastatic LNs, (a) and (b), are quite different in appearance, showing the large intra-class difference. On the right, a normal LN (c) and a metastatic LN (d) appear quite similar, showing the small inter-class difference.

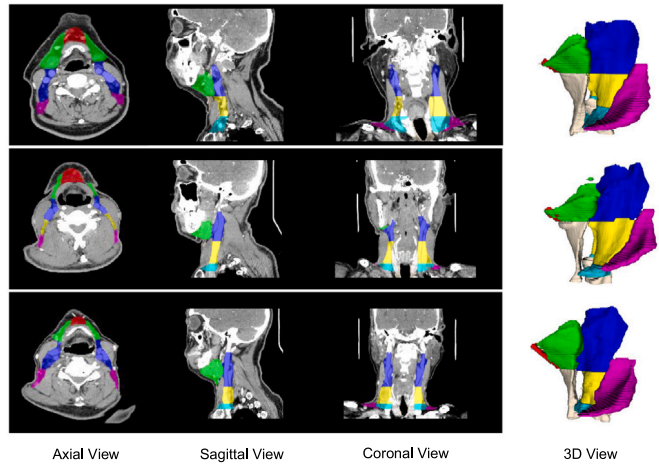

**Fig. 3.** Examples of LN stations. Each row is a patient and the LN stations are marked in different colors.

station is a marking system that describes the area where LNs are sited, usually with edges of various physiological structures (such as muscles and blood vessels) as the boundary (Guo et al., 2021), as shown in Fig. 3. LNs in the same LN station share similar physiological properties. A metastatic LN indicates the other LNs in the same LN station are much more likely to be metastatic. Accordingly, experienced radiologists would focus more on one LN station if they find a metastatic LN in it to improve diagnosis accuracy. In addition, LN stations are connected by lymphatic drainage directly. Therefore, there are strong correlations between neighboring LN stations. If an LN station has been confirmed to contain metastatic LN, the neighboring LN stations are considered more vulnerable.

Compared with other additional information, such as the PET modality and tumor locations, LN stations reflect the metastatic pattern more directly. They can be auto-delineated precisely because their locations are stable, making them more suitable for metastatic LN detection. An LN station segmentation network is trained with a small set of LN segmentation data to locate the LN stations automatically.

Fig. 4 presents the overall framework of our method, where an additional LN station classification branch is introduced to the common detection branch. For the *LN station classification branch*, a new task is introduced to discriminate whether an LN station contains metastatic LNs. To model the interaction between LN stations, a Graph Convolution Network (GCN)-based structure is integrated into the LN station feature encoder. The features extracted this way, on the one hand, can indicate which stations are riskier and help the network focus more on the LNs in these stations so that metastatic LNs will not be missed due to small sizes. On the other hand, the relationship information learned by this structure can supplement the visual features and tackle the problem of ambiguous appearance. For the *metastatic LN detection branch*, following the methods in Zhu et al. (2020b, 2019), we train a metastatic LN candidate generator at high sensitivity. There are massive false-positive cases in these candidates because of the sparsity of metastatic LNs. We feed the LN station features embedding extracted from the LN station classification branch together with the LN candidate into the LN candidate classifier to decrease the false alarm ratio via leveraging LN station information.

To validate the effectiveness of the proposed method, we collected a dataset containing 114 intravenous contrast-enhanced CT scans of oral squamous cell carcinoma (OSCC) patients. Extensive experiments have shown that the proposed method improves the mean free response operating characteristic (mFROC) by 4.77%, maxF1 by 0.0230, and AUC by 0.0303 compared to the state-of-the-art 3D medical detection method (Roth et al., 2015), demonstrating the effectiveness of our method.

The contribution of the paper can be summarized as follows.

- We propose an efficient deep learning based method to leverage the information of LN stations for metastatic LN detection.
- To learn representations for LN stations, we define a closely related auxiliary task to assist metastatic LN detection by discriminating an LN station is metastatic or not.
- To model the mutual influence among different LN stations, we introduce a GCN-based structure for LN station classification.

## 2. Related work

### 2.1. General object detection in medical images

General object detection is viewed as a classic and typical task to be studied for decades in the computer vision community (Zhao et al., 2019). From the perspective of medical image analysis, it is also promisingly valuable. Many medical issues like lesion detection (Yan et al., 2019) and lung module detection (Teramoto et al., 2016) can all be concluded into this task. The popular detection methods can be divided into two types: end-to-end methods (Baumgartner et al., 2021; Li et al., 2019; Luo et al., 2021; Yan et al., 2018, 2019; Zlocha et al., 2019) and two-stage methods (Sun et al., 2019; Ding et al., 2017; Duan et al., 2019; Zhu et al., 2020c). End-to-end methods prefer to dealing the detection task as a whole complete flowchart. These methods have been widely used in universal detection tasks due to their few hyper-parameters and guaranteed performance. nnDetection (Baumgartner et al., 2021) is one of the latest and most representative end-to-end detection methods. It is self-configuring and can achieve encouraging performance in various medical detection tasks.

Two-stage methods decouple the detection task into two sub-tasks: proposal extractor and FP reducing stage. In the first proposal extractor stage, the generator will give many coarse candidate proposals with high sensitivity and high false alarm ratio. The second stage is trained to classify the candidate proposals and dismiss the false positive cases while maintaining an acceptably high sensitivity score. In many sparse-distributed and small object detection task, like lung module detection (Teramoto et al., 2016) and metastatic LN detection task (Zhu et al., 2020c), two-stage methods are proved to be more suitable compared with end-to-end methods (Chao et al., 2020; Zhu et al., 2020a). By separating the raw task into two easier parts, more effective strategies can be designed for different parts. We adopt this design paradigm in our task as well and prove the necessity by comparing our method with the nnDetection (Baumgartner et al., 2021).

*2.2. Metastatic lymph node detection*

Many former works focus on detecting the enlarged LN whose short axis is greater than 10 mm instead of metastatic LNs (Bouget et al., 2019; Barbu et al., 2011; Nogues et al., 2016; Oda et al., 2018) because the detection of enlarged LNs is much easier and enlarged LNs are considered to be highly likely metastatic medically. Conventional detection algorithm (Barbu et al., 2011; Feulner et al., 2013; Kitasaka et al., 2007) based on morphological prior knowledge will design extraction of the visual features carefully. Recent works based on deep learning (Nogues et al., 2016; Oda et al., 2018) bring more possibilities to this field, improving the performance impressively. The success encourages the community to consider the detection of metastatic LNs regardless of their sizes, which is more practical and challenging. Some works have made successful attempts on this task (Chao et al., 2020; Zhu et al., 2020a,c). Zhu et al. (2020a,c) divide the network into two branches to extract features separately via leveraging the distance between LN candidates and the primary tumor. Chao et al. (2020) models the relationship between LN candidates and uses the information of the primary tumor as well. PET modality is also proved to be useful. However, the locations of the primary tumor for the patients after operation is often unknown. PET modality is also costly to collect and in many practical cases, the patients lack the PET/CT for radiologists to refer. Previous methods will fail when these two important information are missing. Therefore, we hope to use the information of LN station to help the detection of metastatic LNs as another optional auxiliary information, so that a good detection accuracy can still be obtained when the above information is missing. LN stations have been proved to be very important reference in the detection of metastatic LNs clinically (Grégoire et al., 2003; Naruke et al., 1978), but, as far as we know, no prior work has considered how to add this important prior knowledge to the automatic detection algorithm. We are the first to leverage the information of LN stations assisting the detection task.

*2.3. Graph convolution network*

Graph neural network (GNN) is first proposed in Scarselli et al. (2008). It combines the strong expressive power of the graph structure with the neural network (Jie et al., 2020). It can be used to model the relationship of the non-Euclidean structure data in many areas like social network (Jin et al., 2019) and protein–protein interaction networks (Fout, 2017). In Kipf and Welling (2016), graph convolution network (GCN) is first proposed by adding the convolution operation into the original GNN structure. GCN has received a lot of attention due to its simplicity, easy understanding, and strong performance (He et al., 2020; Lu et al., 2021).

In medical image analysis, GCN has great potential to be applied in different tasks since it can directly model the relationship between different physiological tissues which cannot be modeled by the regular convolution neural network (CNN) (Chao et al., 2020; Kazmierski and Haibe-Kains, 2021; Mao et al., 2019; Zhao et al., 2020). Kazmierski and Haibe-Kains (2021), Zhao et al. (2020) both use the GCN to combine the information from the different volumes of interest (VOI). When processing images in medical images, images are often sliced and this structure can integrate global information well. Chao et al. (2020) inspires our work a lot. It builds the graph on the LN candidate level and assumes the metastatic LNs may affect each other related to their physical distance. However, we build the graph on the LN station level instead. We argue that due to the uncertainty of the individual, the interaction between the LNs is not so accurate especially when LN candidates have strong sparsity, and building the relationship on a general level will be more suitable.

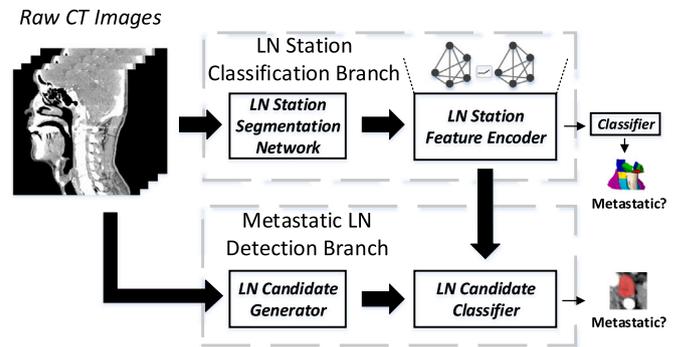

**Fig. 4.** Flowchart of the proposed method, which consists of the LN station classification branch and Metastatic LN Detection branch. The former is leveraged to encode useful features of the LN stations and the latter combines different types of information to make the final decision.

## 3. Method

LN stations play an important role in identifying metastatic LNs. LNs in the same LN station share similar locations and properties. Due to the similarity of LNs in a certain LN station, metastatic LNs usually appear like a cluster. Certain LN stations are more likely to contain metastatic LNs, *e.g.,* station Ib and station II have much more metastatic LNs than others (Fig. 5(a)). There is also a strong correlation between LN stations. An abnormal LN station may reflect that other LN stations near it have more potential to contain metastatic LNs as metastasis may spread to farther LN stations from this abnormal one. Fig. 5(b) shows the correlations between LN stations in terms of metastatic LNs. Some LN stations, such as II_R and III_R, demonstrate strong correlation. As a result, in clinical practice, experienced radiologists typically inspect LN stations first in searching for the metastatic LNs.

Considering the importance of LN stations, we here explore to incorporate information about LN stations into the detection of metastatic LNs, by introducing an additional LN station classification branch to the standard detection branch, as shown in Fig. 4. The features from the two branches are integrated for the final LN candidate classification.

*3.1. Ln station classification branch*

To learn representations for LN stations, the LN station classification branch is trained with a closely related auxiliary task, *i.e.,* to discriminate whether an LN station has metastatic LNs. If an LN station region contains the center point of any metastasis LN, it is considered to have metastatic LNs; otherwise, it is considered not to have metastatic LNs. In this way, the ground truth annotations for the LN station classification can be directly inferred from the annotations of metastasis LNs. With the above task setup, the goal of the LN station classification branch is to learn representations that can best discriminate the LN stations with and without metastasis LNs. To model the relationships among LN stations, a GCN-based classification network is adopted by the LN station classification branch. The overall architecture of the classification network is shown in Fig. 6.

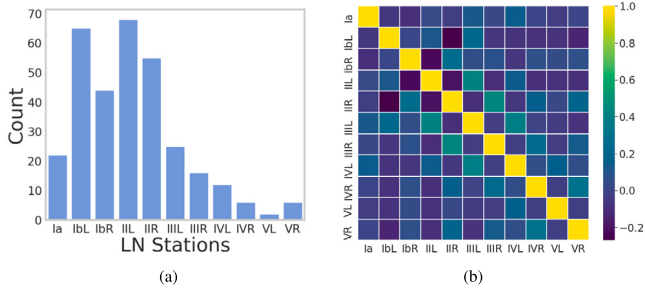

**Fig. 5.** Statistics of LN stations in the case of OSSC. 'L' and 'R' represent the left and right independent parts of the LN station respectively. (a) Statistics of the metastatic LNs contained in each LN station. (b) Spearman's correlation of LN stations regarding to metastatic LNs.

#### 3.1.1. Data preparation

Let $D = \{x_1, x_2, \ldots, x_N\}$ represents a set of 3D CT volumes, where $x_i \in R^{H \times W \times D}$, the subscript $i$ stands for the $i$th patient, and $H$, $W$, $D$ are the height, width, and depth of the 3D CT volumes, respectively. To get the LN station segmentation masks, an nnU-Net (Isensee et al., 2018) based segmentation network is trained on a small dataset of coarsely annotated LN stations.[2] Without loss of generality, we consider a specific 3D CT volume $x$ to simplify the notations. Denote the LN station region masks obtained by the segmentation network as $S = Seg^S(x)$. Based on the segmentation result, the raw CT volume is divided into 12 LN station patches. Noted that since some LN stations, like II station and VI station, are symmetrical and independent with each other, we divide them into left and right two parts based on the central axis and treat them separately. Therefore though we only have 7 station classes in segmentation results, 12 LN station patches are obtained after post-processing, denoted as $S = \{s_1, s_2, \ldots, s_{12}\}$.

#### 3.1.2. Gcn-based classification network

To model the complex reflux relationship among LN stations, a GCN-based classification network is adopted, as shown in Fig. 6. A graph is formed where each LN station is treated as a vertex in a fully connected graph with self-connections. A shared 3D CNN encoder $C^S(\cdot)$ is first used to get the feature embedding for each LN station separately.

$$h_j = C^S(s_j), \quad j \in \{1, \ldots, 12\}. \tag{1}$$

Here $h_j$ represents the encoder output of the $j$th LN stations, which is further fed into a two-layer graph neural network.

$$f_j^S = \sigma_2(relu(\sigma_1(h_j))), \quad j \in \{1, \ldots, 12\}$$
$$\sigma_l(h_j) = \Sigma_{k \in \{1,\ldots,12\}} a_{kj} \phi(h_k), \quad l \in \{1, 2\}, \tag{2}$$

where $\sigma_1(\cdot)$ and $\sigma_2(\cdot)$ are the two layers of the graph neural network, $a_{kj}$ is the edge weight between the $k$th station and the $j$th station, and $\phi$ is a learnable function. In the common setting of GCN, $a \in [0, 1]$ is set based on some similarity metric before the operation of GCN. While in our case, it is hard to define a reasonable similarity between LN stations due to the complexity of the human circulatory system. Therefore we also set $a$ as a learnable parameter so that the network can learn to model the reflux relationship automatically.

For each LN stations, a shared 2-layer MLP is used to get the final classification results:

$$p_j = MLP(f_j^S), \quad j \in \{1, \ldots, 12\}, \tag{3}$$

where $p_j$ is the final prediction result of the $j$th LN station.

---

[2] The traditional registration methods (Avants et al., 2009) can also be used here to obtain the segmentation of the LN station, but the segmentation results are relatively poor. In order not to affect the effect of our subsequent methods, we choose to use the neural network for its better performance.

Considering that metastatic LNs are quite sparse, which causes severe class imbalance problem, focal loss (Lin et al., 2017) is adopted as the training loss:

$$L^S = \begin{cases} -\alpha(1-p_j)^\gamma log(p_j) & y_j = 1 \\ -(1-\alpha)(p_j)^\gamma log(1-p_j) & y_j = 0 \end{cases}, \tag{4}$$

where $\alpha$ and $\gamma$ are hyper-parameters, and $y_j \in \{0, 1\}$ is the ground truth with 1 indicating metastasis and 0 otherwise.

### 3.2. Metastatic LN detection branch

We follow typical two-stage detection methods and divide this task into two stages: extracting LN candidates and classifying the LN candidates.

#### 3.2.1. LN candidate generation

Following the common setting of LN detection tasks (Chao et al., 2020; Zhu et al., 2020c), we employ an LN candidate generator with high sensitivity to generate LN candidates. Specifically, an nnU-Net (Isensee et al., 2018) based segmentation network is trained with pixel-wise metastatic LN annotations. The segmentation results of the network are leveraged to generate LN candidates. We then locate the center point of each connected component and crop out a cube of the size $48 \times 48 \times 48$ centering around the center point as LN candidates. The cube size is expected to be large enough to cover all LN candidates. This coarse result is expected to lead to a high sensitivity score but severely suffer from the problem of false alarms as shown by previous studies (Chao et al., 2020; Roth et al., 2015). Therefore, in the next part, we mainly focus on classifying the metastatic LN candidates to reduce the false alarm ratio.

#### 3.2.2. LN candidate classification

After getting the LN candidates, the goal of the LN candidate classifier is to judge whether an LN candidate is metastatic or not. The network architecture is shown in Fig. 7. Without loss of generality, we consider a specific LN candidate $c \in R^{48 \times 48 \times 48}$. First, a simple 3D CNN network $C^L$ is used to get the visual information of the LN candidate:

$$f^L = C^L(c). \tag{5}$$

The appearance features $f^L$ are expected to provide visual details of the LN candidate patches, such as morphology, texture, and intensity.

In addition to the appearance features of the LN candidate $c$, we try to integrate its corresponding LN station features in classifying the LN candidates to reduce false alarms. The LN station features contain the global information indicating whether there are metastatic LNs in this LN station. In detail, we calculate the Euclidean distance between the LN candidate $c$ and the center of the 12 LN stations and find the nearest LN station.

$$k = \arg \min_{j \in \{1, \ldots, 12\}} d(c, s_j). \tag{6}$$

The $k$th station is thus considered where the LN candidate $c$ belongs to and its feature embedding $f_k^S$ is then concatenated to the visual feature embedding $f^L$.

Some identified LN candidates might not belong to any LN station regions, and the LN station features shall not be used in this case, so we add distance features to prompt the network further when to focus on the LN stations features. Specifically, the distance to each LN station is represented as 12-dimensional features $f^D$, which is further concatenated into the above features. Adding the distance features can also help the network reject the LN candidates that appear in places far away from LN stations, such as the top of the head, which are impossible to contain LNs. The final features are expressed as follow:

$$F = f^L \| f_k^S \| f^D, \tag{7}$$

where $\|$ denotes the concatenation operation.

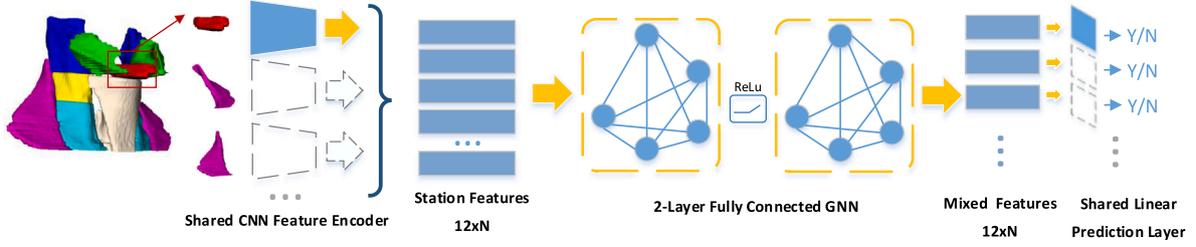

**Fig. 6.** Architecture of the metastatic LN station classification network, consisting of a 3D CNN encoder, a 2-layer GCN, and a 2-layer MLP.

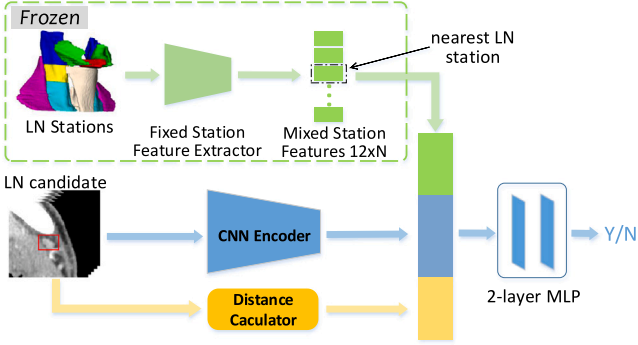

**Fig. 7.** Overall framework of the proposed metastatic LNs classification method. For each LN candidate, a CNN encoder is used to extract its appearance features, which is further integrated with the distance features (marked in yellow) and the features of its nearest LN station (marked in green) before feeding into the MLP for final classification.

Finally we use a 2-layer MLP to get the final classification results.

$$o = MLP(F). \quad (8)$$

The metastatic LN classification task faces the class imbalance problem due to the sparsity of metastatic LNs. The focal loss (Lin et al., 2017) is employed as the training loss:

$$L^L = \begin{cases} -\beta(1-o)^\theta log(o) & y^L = 1 \\ -(1-\beta)o^\theta log(1-o) & y^L = 0 \end{cases}, \quad (9)$$

where $\beta$ and $\theta$ are hyper-parameters, and $y^L \in \{0,1\}$ is the ground truth label of the LN candidate, with 1 representing the metastatic LN candidate and 0 otherwise.

### 3.3. Discussion

It is worth emphasizing that although we use additional LN station annotations information in the training progress, in inference time we do not need extra LN station annotations because the LN station segmentation network automatically generates masks for each LN station. Our method can improve the detection of metastatic LNs only if the patient can provide contrast-enhanced CT images. This makes our solution more in line with actual needs.

## 4. Experiments

### 4.1. Dataset

#### 4.1.1. Lymph node station segmentation

To train the LN station segmentation model, a pixel-level labeled dataset is collected in advance. The dataset contains 82 head & neck intravenous contrast-enhanced CT scans. For each CT scan, the 3D segmentation masks of 7 LN stations of the head & neck are provided by radiologists. To verify segmentation results, we randomly split this dataset into 60%, 20%, 20% for training, validation, and testing respectively.

#### 4.1.2. Metastatic lymph nodes detection

We collected 114 intravenous contrast-enhanced CTs of oral OSCC patients. The median age of these patients is 61 years old and interquartile range is $50 \sim 70$. There are 80 males and 34 females. Patients in this dataset are all under radiotherapy treatments, which means some patients may have undergone surgery, and some parts of their head & neck such as tumors have been manually processed. The metastatic LNs of these patients have been pathologically confirmed. All CT scans are with a slice thickness of 0.625 mm. For evaluation, we randomly split the 114 CT scans into 80 for training and 34 for testing.

### 4.2. Implementation details

In our experiments, the Hounsfield unit value of the CT is clipped to be within $[-218, 600]$. We implement the whole framework stage-by-stage.

#### 4.2.1. LN station segmentation and LN candidate segmentation

A U-Net (Ronneberger et al., 2015) based model is trained to automatically segment the lymph node stations (7 classes) and LN candidates (1 class), respectively. Patches with a shape of $208 \times 238 \times 196$ are randomly cropped to be trained. We further apply random scaling, random gamma, and random mirroring to augment the training data. The 3d U-Net is trained on one GeForce RTX 3090 GPU with a batch size of 2 for 200 epochs, each epoch containing 250 batches. The SGD optimizer with a learning rate of 0.0003 is used with a momentum of 0.99 and a weight decay of 0.00005.

#### 4.2.2. LN station classification

Based on the LN station segmentation masks, we clip the bounding box of each LN station and resize it to $96 \times 96 \times 96$. As some LN stations are symmetrical and independent of each other, we divide them into left and right two parts based on the central axis and treat them separately. Thus, we can get 12 cubes with a size of $96 \times 96 \times 96$ for one CT scan. We treat the 12 cubes as 12 nodes in one graph and use ResNet10 (He et al., 2016) as our feature extractor. 2 graph convolution blocks containing graph convolution layer and Relu activation function are used to exploit the mutual relationship. A fully connected layer is inserted to make the final decision.

#### 4.2.3. LN candidate classification

Every connected component obtained by LN candidate segmentation is treated as an LN candidate and the connected region whose *iou* with the ground truth mask greater than 0.1 is considered as a metastatic case. The center point of each connected component is located, centered around which a patch of $48 \times 48 \times 48$ is cropped.

Both LN station classification and LN candidate classification use the same training setup. Random affine and random elastic deformation are applied in this stage. We trained the model with a batch size of 128 for 100 epochs on one GeForce RTX 3090 GPU. The SGD optimizer with a learning rate of 0.01 is used with a momentum of 0.9 and a weight decay of 0.001. The learning rate changes to 1/10 of the original when the epoch reaches 30, 50, and 70. To handle the category imbalance, we supervise our network using the focal loss parameterized by $\alpha = 0.25, \gamma = 2$.

**Table 1**

Comparison between LN station segmentation by radiologists and neural network in terms of dice and center point distance. Segmentation results are collected from 5 radiologists, and those of the most experienced radiologists are used as ground truth. DSC↑ \CPD↓ are reported.

| Annotator | LN station | | | | | | | |
|---|---|---|---|---|---|---|---|---|
| | Ia | Ib | II | III | IV | V | VI[a] | Average |
| Radiologist1 | 69.18\8.31 | 75.90\12.28 | 73.71\6.52 | 76.86\7.84 | 70.31\8.72 | 70.59\9.68 | 39.10\14.30 | 67.95\9.66 |
| Radiologist2 | 65.50\3.40 | 72.78\11.54 | 60.93\11.17 | 80.76\4.35 | 64.61\10.48 | 65.26\18.37 | 49.66\9.16 | 65.64\9.78 |
| Radiologist3 | 75.52\4.88 | 68.92\11.54 | 73.97\4.63 | 75.04\5.50 | 56.44\16.28 | 49.19\32.00 | 31.57\10.17 | 61.52\12.14 |
| Radiologist4 | 74.52\4.71 | 76.78\12.30 | 74.61\6.72 | 71.21\9.19 | 70.53\7.33 | 65.98\13.94 | 56.57\6.70 | 70.03\8.70 |
| Manual Delineation Average | 71.18\5.33 | 73.59\11.92 | 70.81\7.21 | 75.97\6.72 | 65.47\10.70 | 62.76\18.50 | 44.23\10.08 | 66.29\10.06 |
| Neural Network | 82.56\5.23 | 87.28\3.68 | 87.69\3.82 | 84.82\4.92 | 84.51\4.82 | 81.60\8.63 | 75.88\4.88 | 83.48\5.14 |

[a] There is a long contour line near the esophagus in the mask of the station VI. Since it is very narrow, the criteria for whether to outline this line differs greatly between different people. Therefore, the results of humans on the LN station VI are much worse than the results automatic labeled by the machine.

**Table 2**

Comparison with SOTA metastatic ln detection methods. the mean ± standard deviation are reported.

| Method | FROC@1(%) | FROC@2(%) | FROC@3(%) | FROC@4(%) | mFROC(%) ↑ | Max F1↑ | AUC ↑ |
|---|---|---|---|---|---|---|---|
| nnDetection (Baumgartner et al., 2021) | $43.33_{\pm 6.47}$ | $52.78_{\pm 5.05}$ | $58.06_{\pm 5.08}$ | $59.72_{\pm 4.73}$ | $53.47_{\pm 4.97}$ | $0.5220_{\pm 0.370}$ | $0.6166_{\pm 0.0188}$ |
| single-net(3d classifier)[a] (Zhu et al., 2020a) | $33.64_{\pm 2.83}$ | $53.83_{\pm 3.63}$ | $64.49_{\pm 2.13}$ | $74.58_{\pm 3.26}$ | $56.64_{\pm 1.65}$ | $0.5206_{\pm 0.0150}$ | $0.6673_{\pm 0.0082}$ |
| single-net(2.5d classifier) (Roth et al., 2015) | $34.20_{\pm 1.27}$ | $53.64_{\pm 2.18}$ | $68.41_{\pm 1.09}$ | $77.94_{\pm 3.10}$ | $58.55_{\pm 1.50}$ | $0.5337_{\pm 0.0142}$ | $0.6879_{\pm 0.0103}$ |
| LN-level GCN (Chao et al., 2020) | $23.92_{\pm 1.40}$ | $38.69_{\pm 4.49}$ | $50.47_{\pm 6.11}$ | $63.92_{\pm 5.17}$ | $44.24_{\pm 3.02}$ | $0.4968_{\pm 0.0114}$ | $0.5788_{\pm 0.0257}$ |
| ours | $41.12_{\pm 1.74}$ | $60.12_{\pm 2.50}$ | $71.34_{\pm 1.62}$ | $80.69_{\pm 0.40}$ | $63.32_{\pm 1.04}$ | $0.5571_{\pm 0.0051}$ | $0.7182_{\pm 0.0050}$ |

[a] Many clinical works (Koizumi et al., 2020; Li et al., 2020; Zhang and Ren, 2019; Zhou et al., 2019; Zheng et al., 2021) carry out different experiments for different body positions or different image modalities. Due to the differences of datasets, we cannot directly compare with them, but these methods can be unified into a typical 3D CNN structure, so here we use single-net(3d classifier) to uniformly represent the effects of these works.

### 4.3. Evaluation metric

**DSC:** To evaluate the performance of segmentation tasks, we adopt Dice Similarity Coefficient (DSC) as most medical image segmentation tasks. DSC measures the overlap rate of the prediction mask $p$ and ground truth mask $g$.

$$DSC = \frac{2|p \cap g|}{|p| + |g|}. \quad (10)$$

**CPD:** To evaluate the LN station segmentation results, we calculate the center point distance (CPD) between two segmentation masks. CPD measures the distance error between the prediction mask $p$ and ground truth mask $g$. Let $C_i \in R^3$ denotes the center point coordinate of mask $i$. Voxel euclidean distance is adopted here to measure the distance:

$$CPD = \|C_p - C_g\|_2. \quad (11)$$

**mFROC:** To evaluate the performance of metastatic LN detection tasks, we use the free response operating characteristic (FROC) as the metric, which measures the recall against different numbers of FPs allowed per patient. We report the mFROC, i.e., the average sensitivity at 1,2,3,4 FPs per patient. Since most of the previous work (Chao et al., 2020; Zhu et al., 2020a,c) similar to ours used mFROC as the main reporting indicator, our analysis of method performance also focused on this indicator mainly.

**maxF1:** Besides mFROC, we use maxF1 as a supplementary metric for classification tasks. By adjusting the threshold that decides whether an LN candidate is metastatic, an LN candidate with low confidence to be metastatic can be considered as metastatic with the decrease of the threshold, which leads to different F1-scores with the change of the threshold, formulated as:

$$F1 = \frac{2 \times precision \times recall}{precision + recall}.$$
$$maxF1 = \max_{t \in [0,1]} F1_t \quad (12)$$

**AUC:** AUC represents the area under the receiver operating characteristic (ROC) curve, which is a very common indicator in detection and binary classification tasks.

### 4.4. Quantitative results and discussions

#### 4.4.1. Lymph node station segmentation

To show that LN stations can be precisely auto-delineated, we compare the segmentation results between humans and the segmentation network. The quantitative results are shown in Table 1. On contrast-enhanced CT images, it is difficult to find a border that is totally consistent with the definition of LN stations in medical theory, and the delineation of LN stations depends more on the experience of radiologists. Thus, we collect segmentation masks of 5 radiologists and use the mask of the most experienced radiologist as ground truth to evaluate others' performance. As shown in Table 1, the results of neural network are even better than manual delineation and are sufficient for metastatic LN detection task to crop LN station patches for follow-up stages. This shows LN stations are much less costly to collect compared with other additional information. Fig. 8 shows visualization results of different radiologists and the neural network.

#### 4.4.2. Metastatic LN detection

To show the effectiveness of our method for detecting metastatic LNs, we compare with a set of metastatic LN detection methods, including nnDetection (Baumgartner et al., 2021), a self-configuring framework for 3D medical object detection tasks which has been demonstrated effective on several public benchmarks, as well as several methods especially designed for metastatic LNs detection (Chao et al., 2020; Roth et al., 2015; Zhu et al., 2020a), all of which are based on two-stage framework. For a fair comparison, we use the same LN candidate extractor for the later three methods (Chao et al., 2020; Roth et al., 2015; Zhu et al., 2020a) as our method.

Table 2 outlines the quantitative comparison of different methods for metastatic LN detection tasks, demonstrating the effectiveness of our method. With mFROC of 0.6332, maxF1 of 0.5571, and AUC of 0.7182, our method exceeds nnDetection by 9.85% in mFROC, 0.0351 in maxF1 and 0.1016 in AUC. Our method achieves better results in every metric except FROC@1. nnDetection gives higher confidence to apparent metastatic LNs so it gets a better result at FROC@1. But its drawback is that it tends to overlook small and hard candidates so it cannot achieve a high FROC score even when FP rate allowed per patient is large. Our method outperforms it since we adopt a two-stage methodology to handle small and sparse objects.

Our method achieves a gain of 4.77%, 0.0234, and 0.0303 over the second best method (Roth et al., 2015) in terms of mFROC score, maxF1 score, and AUC score respectively. Compared with Roth et al. (2015) and Zhu et al. (2020a), our method introduces supplementary features on LN stations as well as the distance to LN stations, which are both shown to contribute to a more powerful classifier and effectively reduce the false alarm ratio caused by data imbalance. LN-level GCN (Chao et al., 2020), which treats each LN candidate instead of LN station as a vertex in graph, with the distance between the LN candidates as the edge weights, performs worst among the methods. A possible reason for this phenomenon is that the true metastatic LNs are quite sparse and most LN candidates are actually false positives which provide misguided information. So, it is more reasonable to leverage the relationship information on a more general aspect, such as on the LN station level, to exclude the effect of some individual mistakes.

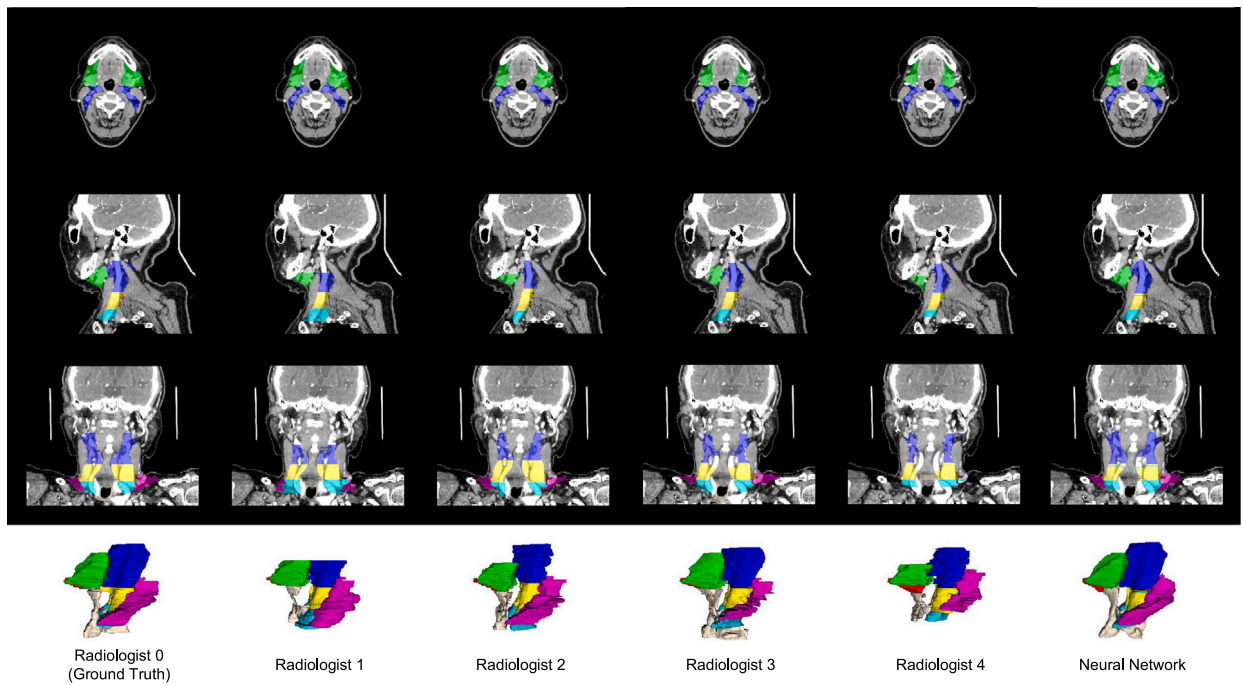

**Fig. 8.** Visualization of LN stations segmentation of a patient. Each column is either from one radiologist or the neural network. The results from radiologists are rough because the masks are delineated slice by slice. The neural network instead generates more smooth results.

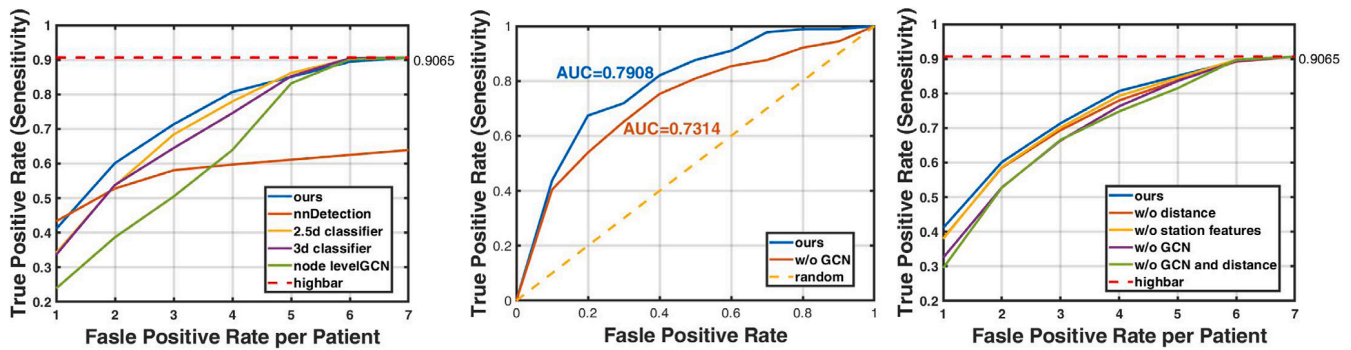

**Fig. 9.** Left: Comparison of our method to SOTA metastatic LNs detection methods in FROC. Middle: ROC of LN station classification with and without GCN. Right: Comparison of the methods in Ablation study in FROC.

Fig. 9 (Left) shows the FROC curves of these methods. There exists an upper bound of 0.9065, because the recall rate for the LN candidate detection is 0.9065 and the missing metastatic LNs cannot be recovered at the classification stage. As a result, the best performance of LN candidate classification cannot exceed this recall rate.

### 4.4.3. Stratified analysis for different LN sizes

In clinical practice, LNs with short axis diameter greater than 10 mm are considered enlarged LNs. While many relevant studies focus on the enlarged LNs (Roth et al., 2015), the majority of metastatic LNs are actually small metastatic LNs. To validate the effectiveness of our method for LNs of different sizes, we separate the metastatic LNs into two groups (*i.e.*, Large and Small) based on their short axis length, using 10 mm as the cut-off threshold. The FROC curves of different methods are separately presented for the above two size groups, as shown in Fig. 11. For all methods, the maxF1, mFROC and AUC scores on the large LNs are better than those on the small LNs, suggesting that it is more challenging to detect metastatic LNs of small sizes. It can be seen that for large LNs, our method outperforms other methods at most sampling points. For small LNs, our method improves the sensitivity by a large margin at different operating points such as 2 FP/patient and 3 FP/patient. Table 3 shows quantitative results of these methods for the two groups. The nnDetection (Baumgartner et al., 2021) performs well for large LNs but terribly on small ones, showing that general detection methods fail to handle the small LNs. For large LNs, our method achieves mFROC of 71.23%, maxF1 of 0.5436 and AUC of 0.8555. For small LNs, our method outperforms the comparative methods by a large margin, *i.e.*, an improvement of 3.72%, 0.0133 and 0.0057 over the second best method, single-net(2.5d classifier), in terms of mFROC, maxF1 and AUC. The above result confirms that introducing the LN station level feature is effective for detecting small metastatic LNs.

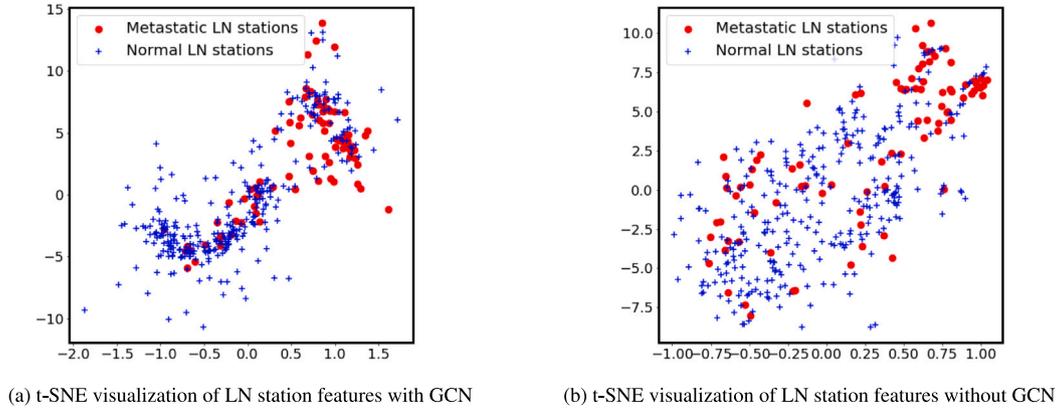

(a) t-SNE visualization of LN station features with GCN

(b) t-SNE visualization of LN station features without GCN

**Fig. 10.** t-SNE visualization of LN station feature distribution in 2D space.

**Table 3**
Improvement on LNs of different sizes. Large lns and small lns are separated based on the short axis (10 mm).

| Method | Large LNs | | | Small LNs | | |
|---|---|---|---|---|---|---|
| | mFROC (%)↑ | maxF1↑ | AUC↑ | mFROC (%)↑ | maxF1↑ | AUC↑ |
| nnDetection (Baumgartner et al., 2021) | 68.86$_{\pm 5.37}$ | 0.5361$_{\pm 0.0112}$ | 0.8304$_{\pm 0.0545}$ | 41.58$_{\pm 4.10}$ | 0.3458$_{\pm 0.0334}$ | 0.5262$_{\pm 0.0347}$ |
| single-net(3d classifier) (Zhu et al., 2020a) | 65.63$_{\pm 0.63}$ | 0.4882$_{\pm 0.0017}$ | 0.7855$_{\pm 0.0096}$ | 46.76$_{\pm 3.06}$ | 0.3657$_{\pm 0.0066}$ | 0.5608$_{\pm 0.0228}$ |
| single-net(2.5d classifier) (Roth et al., 2015) | 65.71$_{\pm 2.28}$ | 0.4832$_{\pm 0.0115}$ | 0.7980$_{\pm 0.0173}$ | 50.69$_{\pm 2.53}$ | 0.3700$_{\pm 0.0099}$ | 0.5887$_{\pm 0.0189}$ |
| LN-level GCN (Chao et al., 2020) | 57.59$_{\pm 1.09}$ | 0.4371$_{\pm 0.0100}$ | 0.7337$_{\pm 0.0149}$ | 30.69$_{\pm 3.83}$ | 0.3370$_{\pm 0.0039}$ | 0.5501$_{\pm 0.0270}$ |
| ours | **71.23**$_{\pm 0.56}$ | **0.5436**$_{\pm 0.0058}$ | **0.8555**$_{\pm 0.0014}$ | **54.41**$_{\pm 1.60}$ | **0.3833**$_{\pm 0.0082}$ | **0.5944**$_{\pm 0.0085}$ |

**Table 4**
Ablation analysis of different components. GCN and w/o GCN represent with and without the gcn-based structure, respectively. distance means distance features. the mean ± standard deviation are reported.

| Method | | | FROC@1(%)↑ | FROC@2(%)↑ | FROC@3(%)↑ | FROC@4(%)↑ | mFROC(%)↑ | maxF1↑ | AUC↑ |
|---|---|---|---|---|---|---|---|---|---|
| Distance | LN station features | | | | | | | | |
| | w/o GCN | GCN | | | | | | | |
| | | | 33.64$_{\pm 2.83}$ | 53.83$_{\pm 3.63}$ | 64.49$_{\pm 2.13}$ | 74.58$_{\pm 3.26}$ | 56.64$_{\pm 1.65}$ | 0.5206$_{\pm 0.0150}$ | 0.6673$_{\pm 0.0082}$ |
| | ✓ | | 29.53$_{\pm 2.99}$ | 52.71$_{\pm 2.10}$ | 66.54$_{\pm 2.53}$ | 74.77$_{\pm 1.18}$ | 55.89$_{\pm 1.74}$ | 0.5194$_{\pm 0.0101}$ | 0.6534$_{\pm 0.0117}$ |
| | | ✓ | 38.13$_{\pm 2.32}$ | 58.50$_{\pm 2.01}$ | 69.35$_{\pm 3.15}$ | 77.94$_{\pm 2.26}$ | 60.98$_{\pm 1.31}$ | 0.5462$_{\pm 0.0160}$ | 0.7060$_{\pm 0.0090}$ |
| ✓ | | | 37.94$_{\pm 4.93}$ | 58.69$_{\pm 2.73}$ | 70.09$_{\pm 1.96}$ | 79.77$_{\pm 1.37}$ | 61.49$_{\pm 0.73}$ | 0.5443$_{\pm 0.0063}$ | 0.7008$_{\pm 0.0118}$ |
| ✓ | ✓ | | 32.52$_{\pm 2.73}$ | 52.81$_{\pm 1.73}$ | 66.36$_{\pm 3.07}$ | 76.26$_{\pm 1.73}$ | 56.96$_{\pm 1.31}$ | 0.5225$_{\pm 0.0097}$ | 0.6668$_{\pm 0.0096}$ |
| ✓ | | ✓ | **41.12**$_{\pm 1.74}$ | **60.12**$_{\pm 2.50}$ | **71.34**$_{\pm 1.62}$ | **80.69**$_{\pm 0.40}$ | **63.32**$_{\pm 1.04}$ | **0.5571**$_{\pm 0.0051}$ | **0.7182**$_{\pm 0.0050}$ |

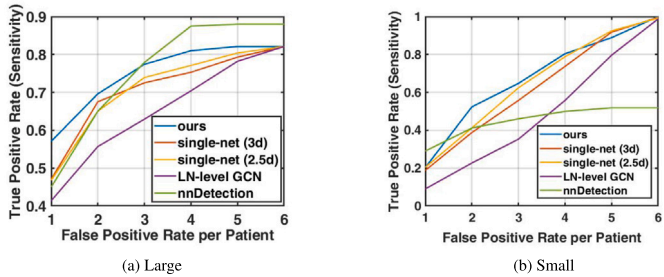

(a) Large

(b) Small

**Fig. 11.** Comparison to SOTA metastatic LNs detection methods in FROC on different size groups. Left: the large metastatic LN group, right: the small one.

### 4.5. Ablation study

For ablation study, we mainly focus on evaluating the effectiveness of the three most significant parts of our method: distance features, LN station features, and GCN structure. Table 4 and Fig. 9 (Right) provide the quantitative comparison among different settings.

#### 4.5.1. Distance features
As can be seen from Table 4, introducing the distance as features leads to clear gains in metastatic LN classification, with 4.84% (line 4 vs. line 1), 1.07% (line 5 vs. line 2), and 2.34% (line 6 vs. line 3) improvement in mFROC score, respectively, suggesting that providing the global spatial position of the LN candidate as distance features is indeed useful. A possible explanation is that metastatic LNs are definitely not possible to be present in certain areas of the head & neck such as the brain, and the distance features make the network capture this kind of knowledge. The distance features, designed to indicate which LN stations are useful for the classification of LN candidates, are also shown to be complementary to the LN station features, as manifested by the gains after involving LN station features.

#### 4.5.2. LN station features
Comparing the results of line 1 with those of line 3 in the Table 4, adding the LN station features can improve the mFROC score by 4.34%, suggesting that it is beneficial to leverage the information on the LN station levels. LN station features express the global visual information of a wide range of related regions, which is designed to imitate the habit of radiologists to search for risky LN stations first. However, modeling the relationship between LN stations is very vital to capture the LN station features. Fig. 9 (Middle) shows that without GCN, the performance of LN station classification drops significantly. We also visualize the LN station features in 2D space using the classical t-SNE method (Statistics, 2011). As shown in Fig. 10, without GCN, the features between metastatic LN stations and normal ones are not clearly separated. Both results suggest that the relationship among LN stations is vital to reflect whether an LN station is metastatic. Consequently, as shown in Table 4, adding the LN station features without GCN even lead to a performance drop (Line 1 vs. Line 2, and Line 4 vs. Line 5).

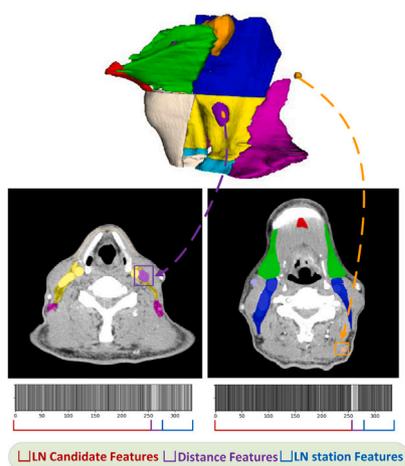

**Fig. 12.** CAM analysis of the result. Light purple represents the positive LN candidate and orange represents the negative LN candidate. The stripes bar on the right is the CAM of the final concatenated features. The first 256-dim features are the LN candidate features. The middle 16-dim features are distance features and the final 64-dim features are the LN station features. The brighter the bar means the greater the impact the features have over the result.

*4.5.3. GCN structure*

To show the effectiveness of the GCN structure, we experiment with ResNet to directly encode the LN station feature (denoted as 'w/o GCN'), *i.e.*, removing the GCN structure in the GCN-based classification network in Fig. 6. We first compare the two methods for LN station classification. As shown in Fig. 9 (Middle), with GCN structure, the method improves the AUC by 0.0594, suggesting that the features extracted by the GCN-based network are much more reliable to indicate whether an LN station is metastatic. We further compare the performance of the two methods on LN classification. As shown in Table 4 (Line 2 vs. Line 3, and Line 5 vs. Line 6), adding the GCN structure always improves the performance of the methods, showing a gain of 6.36% and 5.09%, for with and without distance features, respectively. On the other hand, without the GCN structure, the LN station feature even leads to a performance drop, which also indicates the importance of GCN structure.

*4.6. Class activation map*

To show how different types of features impact the results, we draw the Class Activation Map (CAM) (Zhou et al., 2016) on the final combined features. We follow the algorithm introduced in Zhou et al. (2016) while we calculate heat-map on the final combined features instead of the original 3D images. Fig. 12 provides two examples, where the left shows an LN candidate within the LN station, and the right shows an LN candidate not belonging to any LN station. The CAM bar is put below the images. The abscissa of the CAM bar ranges from 0 to 331. The first 256 dims represent the LN candidate features and the next 12 dims represent the distance features. The final 64 dims are the LN station features. A brighter color on the bar means that the corresponding features have a greater impact on the results. It can be seen that most parts of the CAM map are equally bright for the left case. This shows when a candidate is in an LN station, every type of features is useful for the network and proves the LN station features and distance features are closely related to judging whether LN candidates are metastatic. In the right case, the distance features are much brighter than other parts, indicating the network mainly focuses on the distance features. This shows when the candidate is far away from LN stations (where LNs should not exist), the distance features can be directly used to rule out the candidates. This verifies the importance of keeping distance features to reduce the difficulty of classification tasks and to guide the network when to use other features.

## 5. Conclusion

This paper proposes a new multi-stage metastatic LN detection framework combining the prior information on the LN stations. A new task is defined to identify metastatic LN stations by imitating the behavior of radiologists who examine the risky LN station region first before searching for metastatic LNs. Through solving this task, features of LN stations that are expected to benefit the downstream task, are learned. To model the complex relationship among LN stations, a GCN-based classification model is adopted, which refines the features for LN stations so that they obeys the medical principle that a metastatic LN station affects other LN stations through the circulatory system. Based on the two-stage detection framework, we separate metastatic LN detection task into generating LN candidates and classifying LN candidates. On the classification network, LN station features are combined to decrease the false alarm ratio further. Experiments on a contrast-enhanced CTs dataset of 114 cases with OSSC have shown that our method improves over the current SOTA metastatic LN detection method (Roth et al., 2015) by 4.77%, 0.0234 on the mFROC and maxF1 respectively, demonstrating its effectiveness. Satisfactory performance when handling small metastatic LNs proves the clinical significance of our method.

**CRediT authorship contribution statement**

**Chaoyi Wu:** Methodology, Software, Validation, Writing – original draft. **Feng Chang:** Methodology, Software, Validation, Writing – original draft. **Xiao Su:** Data curation, Investigation. **Zhihan Wu:** Data curation, Investigation. **Yanfeng Wang:** Conceptualization, Writing – review & editing. **Ling Zhu:** Conceptualization, Investigation, Writing – review & editing. **Ya Zhang:** Conceptualization, Methodology, Writing – review & editing.

**Declaration of competing interest**

The authors declare that they have no known competing financial interests or personal relationships that could have appeared to influence the work reported in this paper.

**Acknowledgments**

This work is supported partially by National Key R&D Program of China (No. 2019YFB1804304), SHEITC (No. 2018-RGZN-02046), 111 plan, China (No. BP0719010), and STCSM (No. 18DZ2270700), the "Clinical +Plan" Project of the Shanghai Ninth People's Hospital, Shanghai Jiao Tong University School of Medicine (JYLJ201918) and "Technology transfer and promotion" Project of the Shanghai Jiao Tong University School of Medicine, China (ZT202108).

**References**

Avants, B.B., Tustison, N., Song, G., et al., 2009. Advanced normalization tools (ANTS). Insight J 2 (365), 1–35.
Barbu, A., Suehling, M., Xu, X., Liu, D., Zhou, S.K., Comaniciu, D., 2011. Automatic detection and segmentation of lymph nodes from CT data. IEEE Trans. Med. Imaging 31 (2), 240–250.
Baumgartner, M., Jäger, P.F., Isensee, F., Maier-Hein, K.H., 2021. Nndetection: A self-configuring method for medical object detection. In: de Bruijne, M., Cattin, P.C., Cotin, S., Padoy, N., Speidel, S., Zheng, Y., Essert, C. (Eds.), Medical Image Computing and Computer Assisted Intervention –. MICCAI 2021, Springer International Publishing, Cham, pp. 530–539.
Bouget, D., Jørgensen, A., Kiss, G., Leira, H.O., Langø, T., 2019. Semantic segmentation and detection of mediastinal lymph nodes and anatomical structures in CT data for lung cancer staging. Int. J. Comput. Assist. Radiol. Surg. 14 (6), 977–986.
Chao, C.-H., Zhu, Z., Guo, D., Yan, K., Ho, T.-Y., Cai, J., Harrison, A.P., Ye, X., Xiao, J., Yuille, A., et al., 2020. Lymph node gross tumor volume detection in oncology imaging via relationship learning using graph neural network. In: International Conference on Medical Image Computing and Computer-Assisted Intervention. Springer, pp. 772–782.

Ding, J., Li, A., Hu, Z., Wang, L., 2017. Accurate pulmonary nodule detection in computed tomography images using deep convolutional neural networks. In: International Conference on Medical Image Computing and Computer-Assisted Intervention. Springer, pp. 559–567.

Duan, K., Bai, S., Xie, L., Qi, H., Huang, Q., Tian, Q., 2019. Centernet: Keypoint triplets for object detection. In: Proceedings of the IEEE/CVF International Conference on Computer Vision. pp. 6569–6578.

Feulner, J., Zhou, S.K., Hammon, M., Hornegger, J., Comaniciu, D., 2013. Lymph node detection and segmentation in chest CT data using discriminative learning and a spatial prior. Med. Image Anal. 17 (2), 254–270.

Fout, A.M., 2017. Protein interface prediction using graph convolutional networks. (Ph.D. thesis). Colorado State University.

Grégoire, V., Levendag, P., Ang, K.K., Bernier, J., Braaksma, M., Budach, V., Chao, C., Coche, E., Cooper, J.S., Cosnard, G., et al., 2003. CT-based delineation of lymph node levels and related CTVs in the node-negative neck: DAHANCA, EORTC, GORTEC, NCIC, RTOG consensus guidelines. Radiother. Oncol. 69 (3), 227–236.

Guo, D., Ye, X., Ge, J., Di, X., Lu, L., Huang, L., Xie, G., Xiao, J., Lu, Z., Peng, L., et al., 2021. Deepstationing: thoracic lymph node station parsing in CT scans using anatomical context encoding and key organ auto-search. In: International Conference on Medical Image Computing and Computer-Assisted Intervention. Springer, pp. 3–12.

He, X., Deng, K., Wang, X., Li, Y., Zhang, Y., Wang, M., 2020. Lightgcn: Simplifying and powering graph convolution network for recommendation. In: Proceedings of the 43rd International ACM SIGIR Conference on Research and Development in Information Retrieval. pp. 639–648.

He, K., Zhang, X., Ren, S., Sun, J., 2016. Deep residual learning for image recognition. In: Proceedings of the IEEE Conference on Computer Vision and Pattern Recognition. pp. 770–778.

Isensee, F., Petersen, J., Klein, A., Zimmerer, D., Jaeger, P.F., Kohl, S., Wasserthal, J., Koehler, G., Norajitra, T., Wirkert, S., et al., 2018. Nnu-net: Self-adapting framework for u-net-based medical image segmentation. arXiv preprint arXiv:1809.10486.

Jie, Z.A., Gc, A., Sh, A., Zz, A., Cheng, Y.B., Zl, A., Lw, C., Cl, C., Ms, A., 2020. Graph neural networks: A review of methods and applications. AI Open 1, 57–81.

Jin, D., Liu, Z., Li, W., He, D., Zhang, W., 2019. Graph convolutional networks meet markov random fields: Semi-supervised community detection in attribute networks. In: Proceedings of the AAAI Conference on Artificial Intelligence, Vol. 33. (01), pp. 152–159.

Kazmierski, M., Haibe-Kains, B., 2021. Lymph node graph neural networks for cancer metastasis prediction. arXiv preprint arXiv:2106.01711.

Kipf, T.N., Welling, M., 2016. Semi-supervised classification with graph convolutional networks. arXiv preprint arXiv:1609.02907.

Kitasaka, T., Tsujimura, Y., Nakamura, Y., Mori, K., Suenaga, Y., Ito, M., Nawano, S., 2007. Automated extraction of lymph nodes from 3-D abdominal CT images using 3-D minimum directional difference filter. In: International Conference on Medical Image Computing and Computer-Assisted Intervention. Springer, pp. 336–343.

Koizumi, M., Motegi, K., Koyama, M., Ishiyama, M., Terauchi, T., 2020. Diagnostic performance of a computer-assisted diagnostic system: sensitivity of BONENAVI for bone scintigraphy in patients with disseminated skeletal metastasis is not so high. Ann. Nucl. Med. 34 (3), 200–211.

Li, J., Dong, D., Fang, M., Wang, R., Gao, J., 2020. Dual-energy CT–based deep learning radiomics can improve lymph node metastasis risk prediction for gastric cancer. Eur. Radiol. 30 (4), 2324–2333.

Li, Z., Zhang, S., Zhang, J., Huang, K., Wang, Y., Yu, Y., 2019. Mvp-net: Multi-view fpn with position-aware attention for deep universal lesion detection. In: International Conference on Medical Image Computing and Computer-Assisted Intervention. Springer, pp. 13–21.

Lin, T.-Y., Goyal, P., Girshick, R., He, K., Dollár, P., 2017. Focal loss for dense object detection. In: Proceedings of the IEEE International Conference on Computer Vision. pp. 2980–2988.

Lu, Q., Nguyen, T.H., Dou, D., 2021. Predicting patient readmission risk from medical text via knowledge graph enhanced multiview graph convolution. In: Proceedings of the 44th International ACM SIGIR Conference on Research and Development in Information Retrieval. pp. 1990–1994.

Luo, X., Song, T., Wang, G., Chen, J., Chen, Y., Li, K., Metaxas, D.N., Zhang, S., 2021. SCPM-net: An anchor-free 3D lung nodule detection network using sphere representation and center points matching. arXiv preprint arXiv:2104.05215.

Mao, C., Yao, L., Luo, Y., 2019. MedGCN: Graph convolutional networks for multiple medical tasks.

Naruke, T., Suemasu, K., Ishikawa, S., 1978. Lymph node mapping and curability at various levels of metastasis in resected lung cancer. J. Thorac. Cardiovasc. Surg. 76 (6), 832–839.

Nogues, I., Lu, L., Wang, X., Roth, H., Bertasius, G., Lay, N., Shi, J., Tsehay, Y., Summers, R.M., 2016. Automatic lymph node cluster segmentation using holistically-nested neural networks and structured optimization in CT images. In: International Conference on Medical Image Computing and Computer-Assisted Intervention. Springer, pp. 388–397.

Oda, H., Roth, H.R., Bhatia, K.K., Oda, M., Kitasaka, T., Iwano, S., Homma, H., Takabatake, H., Mori, M., Natori, H., et al., 2018. Dense volumetric detection and segmentation of mediastinal lymph nodes in chest CT images. In: Medical Imaging 2018: Computer-Aided Diagnosis, Vol. 10575. International Society for Optics and Photonics, 1057502.

Ronneberger, O., Fischer, P., Brox, T., 2015. U-net: Convolutional networks for biomedical image segmentation. In: International Conference on Medical Image Computing and Computer-Assisted Intervention. Springer, pp. 234–241.

Roth, H.R., Lu, L., Liu, J., Yao, J., Seff, A., Cherry, K., Kim, L., Summers, R.M., 2015. Improving computer-aided detection using convolutional neural networks and random view aggregation. IEEE Trans. Med. Imaging 35 (5), 1170–1181.

Scarselli, F., Gori, M., Tsoi, A.C., Hagenbuchner, M., Monfardini, G., 2008. The graph neural network model. IEEE Trans. Neural Netw. 20 (1), 61–80.

Statistics, C., 2011. Visualizing data using t-SNE laurens.

Sun, K., Xiao, B., Liu, D., Wang, J., 2019. Deep high-resolution representation learning for human pose estimation. In: Proceedings of the IEEE/CVF Conference on Computer Vision and Pattern Recognition. pp. 5693–5703.

Teramoto, A., Fujita, H., Yamamuro, O., Tamaki, T., 2016. Automated detection of pulmonary nodules in PET/CT images: Ensemble false-positive reduction using a convolutional neural network technique. Med. Phys. 43 (6Part1), 2821–2827.

Yan, K., Bagheri, M., Summers, R.M., 2018. 3D context enhanced region-based convolutional neural network for end-to-end lesion detection. In: International Conference on Medical Image Computing and Computer-Assisted Intervention. Springer, pp. 511–519.

Yan, K., Tang, Y., Peng, Y., Sandfort, V., Bagheri, M., Lu, Z., Summers, R.M., 2019. Mulan: Multitask universal lesion analysis network for joint lesion detection, tagging, and segmentation. In: International Conference on Medical Image Computing and Computer-Assisted Intervention. Springer, pp. 194–202.

Zhang, L., Ren, Z., 2019. Comparison of CT and MRI images for the prediction of soft-tissue sarcoma grading and lung metastasis via a convolutional neural networks model. Clin. Radiol. 75 (1).

Zhao, Y., Yang, F., Fang, Y., Liu, H., Zhou, N., Zhang, J., Sun, J., Yang, S., Menze, B., Fan, X., et al., 2020. Predicting lymph node metastasis using histopathological images based on multiple instance learning with deep graph convolution. In: Proceedings of the IEEE/CVF Conference on Computer Vision and Pattern Recognition. pp. 4837–4846.

Zhao, Z.-Q., Zheng, P., Xu, S.-t., Wu, X., 2019. Object detection with deep learning: A review. IEEE Trans. Neural Netw. Learn. Syst. 30 (11), 3212–3232.

Zheng, Q., Yang, L., Zeng, B., Li, J., Liao, G., 2021. Artificial intelligence performance in detecting tumor metastasis from medical radiology imaging: A systematic review and meta-analysis.

Zhou, B., Khosla, A., Lapedriza, A., Oliva, A., Torralba, A., 2016. Learning deep features for discriminative localization. In: CVPR.

Zhou, L.Q., Wu, X.L., Huang, S.Y., Wu, G.G., Dietrich, C.F., 2019. Lymph node metastasis prediction from primary breast cancer US images using deep learning. Radiology 294 (1), 190372.

Zhu, J., Jin, D., Yan, K., Ho, T.-Y., Ye, X., Guo, D., Chao, C.-H., Xiao, J., Yuille, A., Lu, L., 2020a. Lymph node gross tumor volume detection and segmentation via distance-based gating using 3d ct/pet imaging in radiotherapy. In: International Conference on Medical Image Computing and Computer-Assisted Intervention. Springer, pp. 753–762.

Zhu, Z., Lu, Y., Shen, W., Fishman, E.K., Yuille, A.L., 2020b. Segmentation for classification of screening pancreatic neuroendocrine tumors. arXiv preprint arXiv:2004.02021.

Zhu, Z., Xia, Y., Xie, L., Fishman, E.K., Yuille, A.L., 2019. Multi-scale coarse-to-fine segmentation for screening pancreatic ductal adenocarcinoma. In: International Conference on Medical Image Computing and Computer-Assisted Intervention. Springer, pp. 3–12.

Zhu, Z., Yan, K., Jin, D., Cai, J., Ho, T.-Y., Harrison, A.P., Guo, D., Chao, C.-H., Ye, X., Xiao, J., et al., 2020c. Detecting scatteredly-distributed, small, andcritically important objects in 3d oncologyimaging via decision stratification. arXiv preprint arXiv:2005.13705.

Zlocha, M., Dou, Q., Glocker, B., 2019. Improving RetinaNet for CT lesion detection with dense masks from weak RECIST labels. In: International Conference on Medical Image Computing and Computer-Assisted Intervention. Springer, pp. 402–410.